\documentclass[pra,aps,superscriptaddress,twocolumn,showpacs]{revtex4}%

\usepackage{graphicx}
\usepackage{latexsym}
\usepackage{amsmath}

\begin{document}

\newcommand{\ket} [1] {\vert #1 \rangle}
\newcommand{\bra} [1] {\langle #1 \vert}
\newcommand{\braket}[2]{\langle #1 | #2 \rangle}
\newcommand{\proj}[1]{\ket{#1}\bra{#1}}
\newcommand{\mean}[1]{\langle #1 \rangle}
\newcommand{\opnorm}[1]{|\!|\!|#1|\!|\!|_2}

%\title{Majorization relations in a two-mode squeezer: a possible way towards solving
%the Gaussian minimum entropy conjecture for bosonic channels}

\title{Majorization theory approach to the Gaussian channel minimum entropy conjecture}

\author{Ra\'{u}l Garc\'{\i}a-Patr\'{o}n}
\affiliation{Research Laboratory of Electronics, MIT, Cambridge, MA 02139}
\affiliation{Max-Planck Institut f\"ur Quantenoptik, Hans-Kopfermann-Str. 1, D-85748 Garching, Germany}

\author{Carlos Navarrete-Benlloch}
\affiliation{Research Laboratory of Electronics, MIT, Cambridge, MA 02139}
\affiliation{Departament d'\`Optica, Universitat de Val\`encia, Dr. Moliner 50, 
46100â Burjassot, Spain}

\author{Seth Lloyd}
\affiliation{Research Laboratory of Electronics, MIT, Cambridge, MA 02139}

\author{Jeffrey H. Shapiro}
\affiliation{Research Laboratory of Electronics, MIT, Cambridge, MA 02139}

\author{Nicolas J. Cerf}
\affiliation{Research Laboratory of Electronics, MIT, Cambridge, MA 02139}
\affiliation{Quantum Information and Communication, Ecole Polytechnique de Bruxelles, CP 165, Universit\'e Libre de Bruxelles, 
1050 Bruxelles, Belgium}

\begin{abstract}
A longstanding open problem in quantum information theory is to find the classical capacity of an optical
communication link, modeled as a Gaussian bosonic channel. It has been conjectured that this capacity 
is achieved by a random coding of coherent states using an isotropic Gaussian distribution in phase space. 
We show that proving a Gaussian minimum entropy conjecture for a quantum-limited amplifier is actually sufficient
to confirm this capacity conjecture, and we provide a strong argument towards this proof by exploiting a connection 
between quantum entanglement and majorization theory.
\end{abstract}

\maketitle

During the 1940s, Shannon developed a mathematical theory of the ultimate limits 
on achievable data transmission rates over a communication channel \cite{Shannon1948},
a work that has been central to the advent of our information era.
Since information is necessarily encoded in a physical system and since quantum mechanics
is currently our best theory of the physical world, it is natural to seek the ultimate limits 
on communication set by quantum mechanics. 
Since the 1970s, scientists started investigating the improvements that quantum technologies
may bring to optical communication systems, see e.g. \cite{Yuen1980,Shapiro1984,Caves1994}. 
Because no proper quantum generalization of Shannon's theory existed at that
time, the usual approach was to compare the performance of different encoding and decoding schemes 
for a given optical channel. This provides lower bounds, but does not give the ultimate capacity
nor the optimal quantum encoding and decoding techniques.

In the 1990s, Holevo, Schumacher and Westmoreland \cite{Holevo1998,Schumacher1997}, 
set the basis for a quantum generalization of Shannon's communication theory. 
Consider a quantum channel $\mathcal{M}$ and  a source $\mathcal{A}=\{p_a,\rho_a\}$ 
of independent and identically distributed (i.i.d.) symbols. For each use of the channel $\mathcal{M}$, Alice sends 
the quantum state $\rho_a$ with probability $p_a$, encoding the letter $a$.
One defines the Holevo information
\begin{equation}
\chi(\mathcal{A},\mathcal{M})=S\left(\mathcal{M}(\rho)\right)
-\sum_ap_aS\left(\mathcal{M}(\rho_a)\right),
\label{eq:Holevo_info}
\end{equation}
where $\rho=\sum_ap_a\rho_a$ and $S(\rho)$ is the von Neumann entropy of
the quantum state $\rho$ \cite{footnote}.
The Holevo information $\chi$ gives the highest achievable communication rate over the channel $\mathcal{M}$ 
for a fixed source $\mathcal{A}$, which may require a collective quantum 
measurement over multiple uses of the channel in order to achieve the optimal decoding operation.
By maximizing Eq.~(\ref{eq:Holevo_info}) over the ensemble of i.i.d. sources $\mathcal{A}$ 
under an energy constraint, we obtain the Holevo capacity
\begin{equation}
C_{H}(\mathcal{M})=\max_\mathcal{A}\chi(\mathcal{A},\mathcal{M}).
\label{eq:single_shot_capacity}
\end{equation}
For some highly symmetric channels, such as the qubit depolarizing channel,
the Holevo capacity actually gives the ultimate channel capacity.
For a long time, it was widely believed that this situation prevails for all channels,
that is, it was assumed that input entanglement could not improve the 
classical communication rate over a quantum channel. However, this was disproved 
in \cite{Hastings2009}, so that the best definition of the classical capacity that we currently 
have requires the regularization
\begin{equation}
 C(\mathcal{M})=\lim_{n\rightarrow\infty}\frac{1}{n}C_{H}(\mathcal{M}^{\otimes n}).
\label{eq:regularized_capacity}
\end{equation}
where $\mathcal{M}^{\otimes n}$ stands for $n$ uses of the channel.
 
An important step towards the elucidation of the classical capacity
of an optical quantum channel was made in \cite{Giovannetti2004a}, 
where the authors showed that $C(\mathcal{M})$ of a pure-loss channel---a good (but idealized) approximation of an optical fiber---is achieved 
by a single-use random coding of coherent states using an isotropic Gaussian distribution.
It had long been conjectured that such an encoding 
achieves  $C(\mathcal{M})$ of the whole class of optical channels called
\textit{phase-insensitive} Gaussian bosonic channels \cite{Giovannetti2004a},
including noisy optical fibers and amplifiers. Actually, proving a slightly stronger result known as
the \textit{minimum output entropy} conjecture, namely that coherent states minimize 
the output entropy of phase-insensitive channels, 
would be sufficient to prove this conjecture on the capacity of such channels \cite{Giovannetti2004b}.
Unfortunately, both conjectures have escaped a proof for all phase-insensitive 
channels but the pure-loss one. 

In this Letter, we attempt to prove the minimum output entropy conjecture
for a single use of an arbitrary phase-insensitive Gaussian bosonic channel $\mathcal{M}$,
which is believed to capture the hard part of the conjecture
for multiple uses of the channel.
We show, using a decomposition of any phase-insensitive channel into
a pure-loss channel and a quantum-limited amplifier, that solving the conjecture 
for a quantum-limited amplifier is sufficient. 
This opens a novel way of attacking the conjecture,
using the Stinespring representation of an amplifier channel as a two-mode squeezer,
and exploiting the connection between entanglement and majorization theory.

\begin{figure}[!t!]
\begin{center}
\includegraphics[width=\columnwidth]{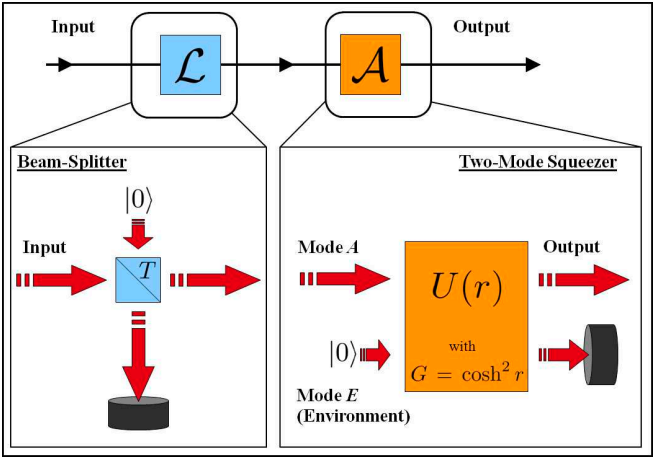}
\end{center}
\caption{Any phase-insensitive Gaussian bosonic channel $\mathcal{M}$ is indistinguishable from a
composed channel $\mathcal{A}\circ\mathcal{L}$, where
$\mathcal{L}$ is a pure-loss channel and $\mathcal{A}$ a quantum-limited amplifier. 
The Stinespring dilation of $\mathcal{L}$ is a beam-splitter of transmissivity $T$,
while the amplifier $\mathcal{A}$ of gain $G$ becomes a two-mode squeezer of parameter $r$ ($G=\cosh^2r$)
in which the input mode $A$ interacts with a vacuum environmental mode $E$.}
\label{Fig:physical_representation}
\end{figure}

\textit{Quantum model of optical channels.-}
A quantum optical channel can be modeled as a Gaussian bosonic channel. 
It is a trace-preserving completely positive map
fully characterized by the action on the Weyl operators
of two $2\times2$ matrices, $K$ and $N$ \cite{Holevo2001,Eisert2007,Weedbrook2011}. 
An intuitive understanding of $K$ and $N$ is given
by the action of the channel on the mean $\bar{x}$ 
and second moments $\gamma$ of the input state,
\begin{equation}
\bar{x}\rightarrow K\bar{x}, \quad   \gamma\rightarrow K\gamma K^T+N.
\end{equation}
For the map to be completely positive, $K$ and $N$
must satisfy \cite{Caruso2004}
\begin{equation}\label{eq:CP_relation}
  N\geq0,\quad \det N\geq(\det K-1)^2.
\end{equation}
Most naturally occurring optical channels, such as optical fibers or amplifiers,
are phase insensitive. 
They correspond to $K={\rm diag}(\sqrt{\tau},\sqrt{\tau})$ and $N={\rm diag}(n,n)$, with
$\tau$ being either the attenuation $0\leq\tau\leq1$ or the amplification $1\leq\tau$ of the channel,
and $n$ being the added noise variance.
Using the composition rule of Gaussian bosonic channels \cite{Caruso2004}, 
it is easy to show that every phase-insensitive channel $\mathcal{M}$ 
is indistinguishable from the concatenation
of a pure-loss channel $\mathcal{L}$ 
of transmissivity $T$ with a quantum-limited amplifier $\mathcal{A}$ 
of gain $G$, see Fig.~\ref{Fig:physical_representation}.
The parameters $T$ and $G$ must satisfy the relations 
$\tau=TG$ and $n=G(1-T)+(G-1)$ in order to guarantee $\mathcal{M}=\mathcal{A}\circ\mathcal{L}$ . 
Three limiting cases are of particular interest: (i) the pure-loss channel, corresponding to 
$G=1$ and $0\leq T\leq1$, having a quantum-limited noise of $n=1-T$;
(ii) the quantum-limited amplifier corresponding to $T=1$ and $G\geq1$, 
with noise $n=G-1$ resulting from spontaneous emission during the amplification process;
(iii) the additive classical noise
channel, corresponding to $\tau=TG=1$ and added thermal noise $n=2G-1$.

\textit{Reduction of the minimum entropy conjecture.-}
As stated earlier, our ultimate goal is to address the following conjecture:
\begin{quote}
\textbf{C1.} Coherent input states minimize the output entropy of any 
phase-insensitive Gaussian bosonic channel $\mathcal{M}$.
\end{quote}
Three simplifications can be made at this point. First,
due to the concavity of the von Neumann entropy, the minimization can be reduced
to the set of pure input states.
Secondly, applying a displacement $D(\alpha)$ at the input of the channel
has the same effect as applying $D(\sqrt{\tau}\alpha)$ at the output, i.e., 
$\mathcal{M}\circ D(\alpha)=D(\sqrt{\tau}\alpha)\circ\mathcal{M}$.  So, because 
the von Neumann entropy is invariant under unitary evolution, 
we can restrict our search to zero-mean input states, that is, 
states $\ket{\varphi}$ satisfying $\bra{\varphi}a\ket{\varphi}=0$ where $a$ is the modal annihilation operator. 
Finally, exploiting the decomposition $\mathcal{M}=\mathcal{A}\circ\mathcal{L}$ 
it is easy to see, using the concavity of the von Neumann entropy, that the minimum output 
entropy of channel $\mathcal{M}$ is lower-bounded by that of channel $\mathcal{A}$, i.e., 
$\mathrm{min}_\phi S(\mathcal{M}(\phi))\geq\mathrm{min}_\psi S(\mathcal{A}(\psi))$ \cite{SupInfo}. 
Since the vacuum state is invariant under $\mathcal{L}$, we conclude
that proving that vacuum minimizes the output  entropy of channel $\mathcal{A}$ implies that 
vacuum also minimizes the output entropy of channel $\mathcal{M}$.

The previous straightforward derivation shows that the conjecture \textbf{C1} 
is strictly equivalent to the following one:

\begin{quote}
\textbf{C2.} Among all zero-mean pure input states, the vacuum state minimizes the 
output entropy of the quantum-limited amplifier $\mathcal{A}$.
\end{quote}

\textit{Entanglement and majorization theory.-}
The Stinespring dilation of a quantum-limited amplifier of gain $G$ is a two-mode squeezer of
parameter $r$, with $G=\cosh^2 r$, which effects the unitary transformation 
(see Fig.~\ref{Fig:physical_representation})
\begin{equation}
 U(r)=\exp\left[r(a_{A}a_{E}-a_{A}^{\dagger}%
a_{E}^{\dagger})/2\right],
\label{eq:Ur} 
\end{equation}
between the input mode $A$  and an environmental mode $E$,
where $a_{X}^{\dagger}$ and $a_{X}$ are the creation and annihilation operators
of mode $X$. 
Because the entanglement $E\left[\ket{\psi}_{AE}\right]$ 
of a pure bipartite state $\ket{\psi}_{AE}$ is uniquely quantified by the von Neumann entropy of its 
reduced density operator $\rho_A={\rm Tr}_E[\ket{\psi}_{AE}\bra{\psi}]$, i.e., 
$E\left[\ket{\psi}_{AE}\right]=S\left(\rho_A\right)$, we can equivalently rephrase 
conjecture \textbf{C2} as
\begin{quote}
\textbf{C3.} Among all input states $\ket{\phi}_{AE}\equiv\ket{\varphi}\otimes\ket{0}$ 
of a two-mode squeezer with $\ket{\varphi}$ having a zero mean, 
the vacuum state $\ket{0}_{AE}\equiv\ket{0}\otimes\ket{0}$ minimizes the output \textit{entanglement}.
\end{quote}

In the remainder of this Letter, we exploit the connection between entanglement and majorization theory 
to attack the proof of \textbf{C3}. Majorization theory provides a partial order relation 
between probability distributions \cite{SupInfo, Arnold}. 
One says that a probability distribution $\mathbf{p}=(p_1,...,p_d)^T$ ($d$ 
might be infinite) majorizes another one $\mathbf{q}$ (denoted $\mathbf{p}\succ \mathbf{q}$) 
if and only if there exists a column-stochastic matrix $D$ (a square matrix whose columns sum to one) 
such that $\mathbf{q}=D\mathbf{p}$, showing that $\mathbf{q}$ is more disordered than $\mathbf{p}$. 
It implies that all concave functions of a distribution, most notably the entropy, can only increase 
along such a ``disorder enhancing" transformation. From an operational point of view, an interesting 
way of proving majorization is by checking the relations
\begin{equation}
\sum_{n=1}^mp_n^{\downarrow}\geq \sum_{n=1}^mq_n^{\downarrow} \quad \forall m<d,
\label{eq:OpMaj}
\end{equation}
where $\mathbf{p}^{\downarrow}$ and $\mathbf{q}^{\downarrow}$ are the original vectors 
with their components rearranged in decreasing order. The notion of majorization can be 
extended to entangled states \cite{Nielsen-Vidal}: a bipartite pure state $\ket{\phi}$ 
majorizes another one $\ket{\psi}$ (noted $ \ket{\phi}\succ \ket{\psi}$) if and only if 
the Schmidt coefficients of $\ket{\phi}$ majorize those of $\ket{\psi}$. This guarantees 
the existence of a deterministic protocol involving only ``local operations and classical 
communication" (LOCC) that maps $\ket{\psi}$ into $\ket{\phi}$, ensuring
the relation $E\left[\ket{\psi}\right] \ge E\left[\ket{\phi}\right]$. We are now ready to introduce
the following stronger conjecture (it implies \textbf{C3}):
\begin{quote}
\textbf{C4.} For any zero-mean state $\ket{\varphi}\neq\ket{0}$, 
the state $U(r)(\ket{\varphi}\otimes\ket{0}$) is majorized 
by the two-mode squeezed vacuum state $U(r)(\ket{0}\otimes\ket{0})$.
\end{quote}

\textit{Infinitesimal two-mode squeezer.-}
Before addressing the general case, let us prove \textbf{C4} 
for an infinitesimal two-mode squeezer by expanding the unitary transformation (\ref{eq:Ur}) 
to the first order in the squeezing parameter $r$,
\begin{equation}
U(r)=I+\frac{r}{2}\left(a_Aa_E-a_A^\dagger a_E^\dagger\right)+O(r^2),
\end{equation}
where $I$ is the identity operator.
Defining the state $\ket{\varphi_\perp}\equiv-a^\dagger_A\ket{\varphi}/(1+\bar{n}_\varphi)^{1/2}$, 
where $\bar{n}_\varphi=\bra{\varphi}a^\dagger_A a_A\ket{\varphi}$ is the mean photon number of the input state $\ket{\varphi}$, 
the output state becomes
\begin{equation}
\ket{\phi_{out}}_{AE}\approx\sqrt{\lambda_\varphi}\ket{\varphi}\otimes\ket{0}
+\sqrt{1-\lambda_\varphi}\ket{\varphi_\perp}\otimes\ket{1},
\label{eq:phi_out}
\end{equation}
with $\lambda_\varphi=1/[1+r^2(\bar{n}_\varphi+1)/4]$.
For any physical state $\ket{\varphi}$ with finite energy $\bar{n}_\varphi$, one can choose
$r$ small enough so that the condition $r \, {\bar{n}_\varphi}^{1/2}\ll 1$ is satisfied
and the approximation (\ref{eq:phi_out}) holds.
%choosing $r$ 
%such that $r\ll 1/\sqrt{\bar{n}_\varphi}$ and $r\ll 1$ are both satisfied.
The key point is to realize that since the input state $\ket{\varphi}$ has a zero mean, 
the states $\ket{\varphi_\perp}$ and $\ket{\varphi}$ are orthogonal, 
so that the state (\ref{eq:phi_out}) is already in Schmidt form. Therefore, 
if $\ket{\varphi}$ and $\ket{\pi}$ are two input states such that $\bar{n}_\varphi<\bar{n}_\pi$, 
then $\lambda_\varphi>\lambda_\pi$,
implying that $U(r)(\ket{\varphi}\otimes\ket{0})\succ U(r)(\ket{\pi}\otimes\ket{0})$
as a result of Eq.(\ref{eq:OpMaj}). 
In other words, any output state is majorized by the states having a lower mean input photon number. 
Finally, since the vacuum state has the minimum mean photon number ($\bar{n}_\varphi=0$), 
this majorization relation proves conjecture \textbf{C4} for infinitesimal two-mode squeezers.

\textit{Majorization relations in a two-mode squeezer.-}
In order to address the conjecture \textbf{C4} for any $r$, let us consider the number-state 
expansion of an arbitrary input state $\ket{\varphi}=\sum_{k=0}^{\infty}c_k\ket{k}$, which leads to the output state
\begin{equation}
U(r)(\ket{\varphi}\otimes\ket{0})=\sum_{k=0}^{\infty}c_k\ket{\Psi^{(k)}_\lambda},
\end{equation}
where $\lambda=\tanh r$ and $\ket{\Psi^{(k)}_\lambda}$ stands for the output state 
corresponding to an input Fock state $\ket{\varphi}=\ket{k}$. As shown in \cite{SupInfo}, we have
\begin{equation}
\ket{\Psi^{(k)}_\lambda}=\sum_{n=0}^\infty\sqrt{p^{(k)}_n(\lambda)}\ket{n+k}\otimes\ket{n},
\label{eq:k_photons}
\end{equation}
with Schmidt coefficients
\begin{equation}
p^{(k)}_n(\lambda)=(1-\lambda^2)^{k+1}\lambda^{2n}\binom{n+k}{n}.
\end{equation}
We have been able to prove two chains of majorization relations by considering either 
different Fock states $\ket{k}$ at the input (for a fixed squeezing parameter $r$) or 
different values of $r$ (for a fixed input Fock state $\ket{k}$). 
First, when restricting to Fock states $\ket{k}$, we can prove that
\begin{equation}
 \ket{\Psi^{(k)}_\lambda} \succ \ket{\Psi^{(k+1)}_\lambda},
\label{eq:majorization_photon_input}
\end{equation}
since there exists a column-stochastic matrix
\begin{equation}
 D_{nm}=(1-\lambda^2)\lambda^{2(n-m)}H(n-m),
\end{equation}
such that $\mathbf{p}^{(k+1)}(\lambda)=D\mathbf{p}^{(k)}(\lambda)$, where $H(x)$ is 
the Heaviside step function defined as $H(x)=0$ for $x<0$ and $H(x)=1$ for $x\geq0$. 
The details of the proof are provided in \cite{SupInfo}, where we also give the explicit 
form of an LOCC protocol that deterministically maps $\ket{\Psi^{(k+1)}_\lambda}$ into 
$\ket{\Psi^{(k)}_\lambda}$. Iterating this procedure, we can easily prove that 
$\ket{\Psi^{(k)}_\lambda}\succ \ket{\Psi^{(k')}_\lambda}$, $\forall k'\geq k$, 
for which we also give the corresponding column-stochastic matrix and deterministic 
LOCC protocol. 

For our matters here, the central consequence is that $\ket{\Psi^{(0)}_\lambda} \succ \ket{\Psi^{(k)}_\lambda}$, $\forall k\geq 0$,
that is, we have proved conjecture \textbf{C4} for the restricted, but complete, set of input Fock states.
Remarkably, this would be sufficient to prove 
the single-use minimum entropy conjecture if it could be shown that the output-entropy minimizing input 
state is isotropic, i.e., its Wigner distribution is rotationally invariant. This is because 
the Fock states are the only isotropic, zero-mean pure states.

\begin{figure}[t]
\begin{center}
\includegraphics[width=\columnwidth]{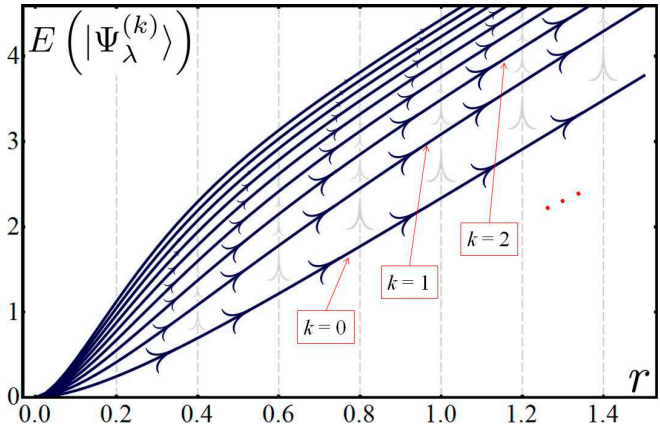}
\end{center}
\caption{Entanglement of the output state $\ket{\Psi^{(k)}_\lambda}$ as a function of the squeezing parameter $r$.
As explained in the text the entanglement is monotonically increasing with $r$ for all Fock input states,
while, for a fixed $r$, it monotonically increases with $k$. This behavior is in full agreement with the majorization
relations (\ref{eq:majorization_photon_input}) and (\ref{eq:majorization_squeezing})
proved in the text. The arrows in the figure indicate the majorization order.}
\label{Fig:Fock_states}
\end{figure}

Second, given an input Fock state $\ket{k}$, one can show that there exists a 
majorization relation in the direction of decreasing squeezing parameter, that is,
\begin{equation}
\ket{\Psi^{(k)}_{\lambda'}}\succ\ket{\Psi^{(k)}_\lambda} \quad \forall\lambda'<\lambda,
\label{eq:majorization_squeezing}
\end{equation}
since one can build \cite{SupInfo} a column-stochastic matrix
\begin{eqnarray}
R^{(k)}_{nm}=&=&\binom{m+k}{m}^{-1}\left(\frac{1-\lambda^2}{1-\lambda'^2}\right)H(n-m)
\\
&&\times\left[L_{n-m}^{(k,m)}\lambda^2-L_{n-m-1}^{(k,m+1)}\lambda'^2\right]\lambda^{2(n-m-1)}, \nonumber
\end{eqnarray}
with
\begin{equation}
L_{m}^{(k,n)}=n\binom{n+k}{k}\binom{m+k}{k} \lambda'^{-2n}B(\lambda'^2;n,1+k),
\end{equation}
and $B(z;a,b)=\int_{0}^{z}dx \; x^{a-1}(1-x)^{b-1}$ being the incomplete beta function, 
such that $\mathbf{p}^{(k)}(\lambda)=R^{(k)}(\lambda,\lambda')\mathbf{p}^{(k)}(\lambda')$. 
In \cite{SupInfo} we give a deterministic LOCC protocol performing the transformation 
$\ket{\Psi^{(k)}_{\lambda}}\rightarrow\ket{\Psi^{(k)}_{\lambda'}}$.

In Fig.\ref{Fig:Fock_states}, we summarize the two chains of majorization relations and their
implications on the output entanglement. From this, as well as the case of the infinitesimal two-mode squeezer, 
it is tempting to conclude that $\bar{n}_\varphi<\bar{n}_\pi$ always implies 
$U(r)(\ket{\varphi}\otimes\ket{0})\succ U(r)(\ket{\pi}\otimes\ket{0})$. 
However, we have numerically  observed that this does not hold in general, 
which probably reflects the difficulty of proving the conjecture. As a concrete example, 
we note that the state $U(r)[(\sqrt{0.4}\ket{1}+\sqrt{0.6}\ket{2})\otimes\ket{0}]$ 
has $\bar{n}=1.6$ mean input photons but is \textit{less} entangled for $r\gtrsim0.75$ 
than $\ket{\Psi^{(1)}_\lambda}$. Nevertheless, our numerical 
investigations have shown that for an arbitrary input state  $\ket{\varphi}$, the 
output states corresponding to different squeezing parameters satisfy the majorization 
relation $U(r')(\ket{\varphi}\otimes\ket{0})\succ U(r)(\ket{\varphi}\otimes\ket{0})$ 
for $r'<r$. Furthermore, we have numerically checked that for a fixed $r$, the majorization 
relation $U(r)(\ket{0}\otimes\ket{0})\succ U(r)(\ket{\varphi}\otimes\ket{0})$
is satisfied by tens of thousands of random superpositions of the first 21 Fock states,
which strongly suggests that conjecture \textbf{C4} holds.

\textit{Conclusion.-}
Using the decomposition of phase-insensitive Gaussian bosonic channels into a pure-loss channel
and a quantum-limited amplifier, we have shown that proving a reduced conjecture for the quantum-limited amplifier
is sufficient to prove the single-use minimum entropy conjecture.
Using Stinespring's theorem, 
this boils down to proving that the vacuum minimizes the output entanglement of a two-mode squeezer.
Then, using the connection between entanglement and majorization theory, 
we have provided a partial proof of this conjecture for a special class of input states, namely photon number states,
as well as a full solution for the infinitesimal channel. To prove the conjecture in general, we are left with
the (possibly simpler) task of showing that the output-entropy minimizing input state is isotropic in phase space, 
that is, no symmetry breaking occurs.
Thus, apart from reinforcing the conjecture even further, we believe that our analysis offers 
a new possible approach to its proof.

The authors would like to thank G. Giedke and J. I. Cirac for helpful discussions. 
C.N.-B. and N.J.C. thank the Optical and Quantum Communications Group at RLE for their hospitality.
R. G.-P., N.J.C., J.H.S., and S.L. acknowledge financial support from the W. M. Keck Foundation Center 
for Extreme Quantum Information Theory, R.G.-P. from the Humboldt foundation, 
C.N.-B. from the FPU program of the MICINN,
J.H.S. and S.L. from the ONR Basic Research Challenge Program,
and N.J.C. from the F.R.S.-FNRS under project HIPERCOM.

\appendix*

\section{Supplementary Information}
In what follows, we give a more complete overview of the calculations leading to the main results of this Letter. 
First, we derive the lower bound used to reduce conjecture \textbf{C1} to \textbf{C2}. Second,
we review the concept of majorization in probability theory, and describe its use in the context of quantum entanglement. 
Then, we detail the calculation of the output state of a two-mode squeezer for an arbitrary input 
state expressed as a superposition of Fock states. Finally, 
we provide a detailed derivation of the chain of majorization relations that are obeyed by a two-mode squeezer 
with number-state inputs in one port, and present their associated local operation and 
classical communication (LOCC) protocols.

\subsection{Reduction of the minimum entropy conjecture}
In what follows we exploit the decomposition $\mathcal{M}=\mathcal{A}\circ\mathcal{L}$ 
and the concavity of the von Neumann entropy to prove that the minimum output 
entropy of channel $\mathcal{M}$ is lower-bounded by that of channel $\mathcal{A}$, i.e., 
$\mathrm{min}_\phi S(\mathcal{M}(\phi))\geq\mathrm{min}_\psi S(\mathcal{A}(\psi))$. 

Let $|\phi\rangle$ be an input pure state of channel $\mathcal{M}$. After
passage through the pure-loss channel $\mathcal{L}$, the intermidiate state 
(between $\mathcal{L}$ and $\mathcal{A}$) is $\tilde{\sigma}=\mathcal{L}(\proj{\phi})$.
For any decomposition
$\{p_i,\psi_i\}$ of $\tilde{\sigma}$ satisfying $\tilde{\sigma}=\sum_ip_i\proj{\psi_i}$,
we have the following chain of inequalities
\begin{eqnarray}
S\left(\mathcal{M}(\proj{\phi})\right)&\stackrel{(1)}{=}&S\left(\mathcal{A}(\tilde{\sigma})\right)
\stackrel{(2)}{=}
S\left(\sum_ip_i\mathcal{A}(\proj{\psi_i})\right) \nonumber\\ 
&&\stackrel{(3)}{\geq} \sum_ip_iS\left(\mathcal{A}(\proj{\psi_i})\right) \nonumber\\ 
&&\stackrel{(4)}{\geq}\mathrm{min}_\psi S(\mathcal{A}(\psi)),
\label{Eq:reduction} 
\end{eqnarray}
%where we used; in (1) the channel decomposition $\mathcal{M}=\mathcal{A}\circ\mathcal{L}$;
%in (2) the linearity of quantum operations; in (3) the sub-additivity of von Neumann entropy; 
%and finally in (4) the definition of the minimum output entropy of channel $\mathcal{A}$.
where we have used: the channel decomposition $\mathcal{M}=\mathcal{A}\circ\mathcal{L}$ in (1); 
the linearity of quantum operations in (2), the sub-additivity of von Neumann entropy in (3);
and, finally, the definition of the minimum output entropy of channel $\mathcal{A}$ in (4).
The proof concludes by noticing that Eq.~(\ref{Eq:reduction}) 
holds for every input state of channel $\mathcal{M}$,
including the one minimizing the output entropy of $\mathcal{M}$.

\subsection{Majorization and Entanglement}

Majorization appeared as a way to order probability distributions in terms of their disorder, 
in an effort to understand when one distribution can be built from another by randomizing the 
later \cite{Arnold}. Take two probability vectors 
$\mathbf{p}=(p_1,p_2,...,p_d)^T$ and $\mathbf{q}=(q_1,q_2,...,q_d)^T$ of dimension $d$ 
(which can be infinite as in our case), properly normalized, that is, $\sum_{n=1}^dp_n=\sum_{n=1}^dq_n=1$. 
We say that $\mathbf{p}$ majorizes $\mathbf{q}$, and denote it by $\mathbf{p}\succ\mathbf{q}$, if and only if
\begin{equation}
 \sum_{n=1}^mp_n^{\downarrow}\geq \sum_{n=1}^mq_n^{\downarrow} \;\; \forall m\le d,
\end{equation}
where $\mathbf{p}^{\downarrow}$ and $\mathbf{q}^{\downarrow}$ are the original vectors with their components 
rearranged in decreasing order. This definition is useful from a practical point of view, 
since it is easy to check numerically if two vectors satisfy these relations. 
Nevertheless, it can be proven that $\mathbf{p}\succ\mathbf{q}$ is strictly equivalent to two other operational relations:
\begin{quote}
 \textbf{M1.} For every concave function $h(x)$, we have $\sum_{n=1}^dh(p_n)\leq\sum_{n=1}^dh(q_n)$.
\end{quote}
\begin{quote}
\textbf{M2.} $\mathbf{q}$ can be obtained from $\mathbf{p}$ via $\mathbf{q}=D\mathbf{p}$, where $D$ is a column-stochastic matrix.
\end{quote}
A square matrix D is column-stochastic if its elements are real and positive, its columns sum to one, 
and its rows sum to less than one. Most of the literature on the connection between 
majorization and quantum information studies finite-dimensional systems, 
in which case it can be shown that column-stochastic matrices are also 
doubly-stochastic (columns and rows both sum to one). 
In this work we need the slightly more general definition of column-stochastic to cope 
with infinite dimensional spaces \cite{Kaftal10}. 
Physically, stochastic matrices are equivalent to convex mixtures of permutations of the vector components,
and hence, property \textbf{M2} shows that $\mathbf{q}$ is more disordered than $\mathbf{p}$.

Interestingly, majorization theory can also be used to answer the question of whether Alice an 
Bob can transform a shared bipartite pure state $\ket{\psi}_{AB}$ into $\ket{\varphi}_{AB}$ 
by using a deterministic protocol involving only local operations and classical communication (LOCC) 
\cite{Nielsen99,Nielsen-Vidal}. Given the probability vectors $\mathbf{p}_{\psi}$ and $\mathbf{p}_{\varphi}$ 
generated with the Schmidt coefficients of these states (the eigenvalues of the reduced density operators), 
it is possible to prove that the transformation $\ket{\psi}_{AB}\rightarrow\ket{\varphi}_{AB}$ 
is possible if and only if $\mathbf{p}_{\varphi}\succ\mathbf{p}_{\psi}$, that is, if the 
Schmidt coefficients of $\ket{\varphi}_{AB}$ majorize those of $\ket{\psi}_{AB}$, 
in which case we use the symbolic notation $\ket{\varphi}_{AB}\succ\ket{\psi}_{AB}$.
The entanglement of a pure bipartite state $\ket{\psi}_{AB}$ being measured by 
the von Neumann entropy of the reduced density operator $\rho_A=\mathrm{Tr}_B[\ket{\psi}_{AB}]$, 
and the von Neumann entropy being a concave function, one gets as an intuitive corollary 
that $\ket{\psi}_{AB}$ can only be transformed deterministically by an 
LOCC protocol into states of lower entanglement, i.e.,
\begin{equation}
 E[\ket{\psi}_{AB}]\geq E[\ket{\varphi}_{AB}],
\end{equation}
as follows from property \textbf{M1}.

Note that while $\ket{\varphi}_{AB}\succ\ket{\psi}_{AB}$ implies that $\mathbf{p}_{\varphi}$ 
can be transformed into $\mathbf{p}_{\psi}$ by application of a column-stochastic matrix, 
the transformation goes in the opposite direction for the corresponding states, 
that is, it is $\ket{\psi}_{AB}$ the state which can be transformed into $\ket{\varphi}_{AB}$ 
by a deterministic LOCC protocol. In other words, at the level of probability 
distributions the transformation induces disorder (increases the entropy), 
while at the level of states the transformation decreases the entanglement, 
as corresponds to physical deterministic LOCC protocols.

\subsection{Output States of a Two-Mode Squeezer}

If we inject the vacuum state at the input of a two-mode squeezer $U(r)$, we obtain the 
two-mode squeezed vacuum state
\begin{equation}
\ket{\Psi^{(0)}}=U(r)\ket{0,0}=\frac{1}{\cosh r}\sum_{n=0}^\infty\tanh^n\hspace{-0.08 cm}r \hspace{0.08 cm} \ket{n,n},
\end{equation}
where $\ket{n}$ is a number state, and we use the compact notation $\ket{m}_A\otimes\ket{n}_B=\ket{m,n}$.

Consider now the more general input state
\begin{equation}
 \ket{\phi}=\ket{\varphi}\otimes\ket{0}=\sum_{n=0}^{\infty}c_n\ket{n,0},
\label{eq:input_state}
\end{equation}
written in the number state basis, which becomes the state
\begin{equation}
\ket{\phi_{out}}=U(r)\ket{\phi}=\sum_{n=0}^{\infty}c_n\ket{\Psi^{(n)}},
\end{equation}
with
\begin{equation}
\ket{\Psi^{(k)}}=U(r)\ket{k,0},
\end{equation}
after passing through the two-mode squeezer.

In the reminder of this section, we focus on finding a manageable expression for the states $\ket{\Psi^{(k)}}$, that is, for the output state of the two-mode squeezer when a number state $\ket{k}$ is fed through one of its input ports. We start by noting that $\ket{\Psi^{(k)}}$ can be written in terms of the two-mode squeezed vacuum state $\ket{\Psi^{(0)}}$ as follows
\begin{equation}
\ket{\Psi^{(k)}}=\frac{1}{\sqrt{k!}}U(r)a^{\dagger k}_A\ket{0,0}=
\frac{1}{\sqrt{k!}}[U(r)a^\dagger_A U(r)^\dagger]^k\ket{\Psi^{(0)}},
\end{equation}
which, using the relation
\begin{equation}
U(r) a^\dagger_A U(r)^\dagger=\cosh r \hspace{0.1 cm} a^\dagger_A-\sinh r \hspace{0.1 cm} a_B,
\end{equation}
can be rewritten as
\begin{equation}
\ket{\Psi^{(k)}}=\sum_{j=0}^k\frac{(-1)^{k-j}}{\sqrt{k!}}\binom{k}{j}\cosh^j\hspace{-0.08 cm}r \sinh^{k-j}\hspace{-0.08 cm}r 
\hspace{0.1 cm} a^{\dagger j}_A a^{k-j}_B \ket{\Psi^{(0)}}.
\label{eq:Kphotons}
\end{equation}
Now, an easy calculation shows that
\begin{eqnarray}
&&a_B \ket{\Psi^{(0)}} = \frac{1}{\cosh r}\sum_{n=1}^\infty\sqrt{n}\tanh^n\hspace{-0.08 cm}r \hspace{0.08 cm} \ket{n,n-1}
\\
&&\underset{n\rightarrow m+1}{=} \frac{1}{\cosh r}\sum_{m=0}^\infty\sqrt{m+1}
\tanh^{m+1}\hspace{-0.08 cm}r \hspace{0.08 cm} \ket{m+1,m}, \nonumber
\end{eqnarray}
leading to the following identity
\begin{equation}
a_B \ket{\Psi^{(0)}} = \tanh r \hspace{0.1 cm} a^\dagger_A \ket{\Psi^{(0)}},
\end{equation}
which allows us to rewrite (\ref{eq:Kphotons}) as
\begin{equation}
\ket{\Psi^{(k)}}=\frac{\cosh^k\hspace{-0.08 cm}r}{\sqrt{k!}}\sum_{j=0}^k(-1)^{k-j}\binom{k}{j} 
\tanh^{2(k-j)}\hspace{-0.08 cm}r \hspace{0.08 cm} a^{\dagger k}_A \ket{\Psi^{(0)}}.
\end{equation}
Finally, using the relations
\begin{subequations}
\begin{eqnarray}
\sum_{j=0}^k (-1)^{k-j} \binom{k}{j} x^{k-j} &=& (1-x)^k,
\\
1-\tanh^2 r &=& \cosh^{-2}r,
\end{eqnarray}
\end{subequations}
we can write the previous expression as
\begin{eqnarray}
\ket{\Psi^{(k)}} &=& \frac{1}{\sqrt{k!}\cosh^k\hspace{-0.08 cm}r} a^{\dagger k}_A \ket{\Psi^{(0)}}
\\
&=& \frac{1}{\cosh^{k+1}r}\sum_{n=0}^\infty \sqrt{\binom{n+k}{k}} \tanh^n \hspace{-0.08 cm}r \hspace{0.08 cm} \ket{n+k,n}. \nonumber
\label{eq:FockOutputState}
\end{eqnarray}
Let us define $\lambda=\tanh r$; from now on we will use the notation
\begin{equation}
\ket{\Psi^{(k)}_\lambda} = \sum_{n=0}^\infty \sqrt{p^{(k)}_n(\lambda)} \ket{n+k,n},
\end{equation}
with
\begin{equation}
p^{(k)}_n(\lambda) = (1-\lambda^2)^{k+1} \lambda^{2n} \binom{n+k}{n},
\end{equation}
to stress the dependence of the state on the squeezing parameter.
Note that the states (\ref{eq:FockOutputState}) are already written in Schmidt form, and in the following we will use 
\begin{equation}
\mathbf{p}^{(k)}= (p^{(k)}_0,p^{(k)}_1,...)^T,
\end{equation}
to denote the corresponding probability vectors.

\subsection{Proof of the Majorization Relations for Fock State Inputs}

In this section we will explain how to derive the column-stochastic matrices needed to prove the majorization relations employed in the Letter.

\subsubsection{Proof of $\ket{\Psi_\lambda^{(k)}}\succ\ket{\Psi_\lambda^{(k+1)}}$}

Because the states $\ket{\Psi_\lambda^{(k)}}$ are already in Schmidt form as commented previously, we need to prove that there exists a column-stochastic matrix $D$ such that
\begin{equation}
 \mathbf{p}^{(k+1)}=D\mathbf{p}^{(k)}.
\end{equation}
This is actually quite simple if one notices that the Pascal identity
\begin{equation}
 \binom{n+k+1}{k+1}=\binom{n+k}{k}+\binom{n+k}{k+1},
\end{equation}
implies the following relation (with the convention $p_n^{(k)}=0$ for $n<0$):
\begin{equation}
 p_n^{(k+1)}=(1-\lambda^2)p_n^{(k)}+\lambda^2p_{n-1}^{(k+1)}.
\end{equation}
This recurrence allows us to connect $\mathbf{p}^{(k+1)}$ with $\mathbf{p}^{(k)}$ by means of a lower-triangular matrix
\begin{equation}
\left( \begin{array}{c}
p_0^{(k+1)} \\
p_1^{(k+1)} \\
p_2^{(k+1)} \\
\vdots
\end{array}\right)
 =(1-\lambda^2)
\left( \begin{array}{cccc}
1 & 0 & 0 &  \ldots \\
\lambda^2 & 1 & 0 &  \ldots \\
\lambda^4 & \lambda^2 & 1 &  \ldots\\
\vdots & \vdots & \vdots & \ddots
\end{array} \right)
\left( \begin{array}{c}
p_0^{(k)} \\
p_1^{(k)} \\
p_2^{(k)} \\
\vdots
\end{array}\right),
\end{equation}
or in a more compact notation
\begin{equation}
 p_n^{(k+1)}=\sum_{m=0}^n(1-\lambda^2)\lambda^{2m}p_{n-m}^{(k)}.
\label{eq:relation1}
\end{equation}
It is fairly easy to show that the triangular matrix shown above, whose elements are explicitly given by
\begin{equation}
 D_{nm}=(1-\lambda^2)\lambda^{2(n-m)}H(n-m),
\end{equation}
with $H(x)$ being the Heaviside step function defined as $H(x)=1$ for $x\geq0$ and $H(x)=0$ for $x<0$, 
is column-stochastic.  Hence we conclude that $\ket{\Psi_\lambda^{(k)}}\succ\ket{\Psi_\lambda^{(k+1)}}$ 
as commented in the Letter.

\subsubsection{Proof of $\ket{\Psi_\lambda^{(k)}}\succ\ket{\Psi_\lambda^{(k+\Delta k)}}$ for $\Delta k>0$}

It is clear that $\ket{\Psi_\lambda^{(k)}}\succ\ket{\Psi_\lambda^{(k+1)}}$ 
implies $\ket{\Psi_\lambda^{(k)}}\succ\ket{\Psi_\lambda^{(k+\Delta k)}}$ 
for all $\Delta k>0$ (note that $\Delta k$ is a positive integer by definition), 
as majorization is clearly a transitive relation. 
This shows that when restricted to Fock-state inputs, 
the output entanglement of a two-mode squeezer increases monotonically with the number of input photons.

In order to find the explicit column-stochastic matrix $D^{(\Delta k)}$ 
satisfying $\mathbf{p}^{(k+\Delta k)}=D^{(\Delta k)}\mathbf{p}^{(k)}$, 
we use the independence on $k$ of the matrix $D$
% used to prove $\mathbf{p}^{(k+1)}=D\mathbf{p}^{(k)}$, 
which allows us 
%to evaluate the needed matrix as
write
\begin{equation}
D^{(\Delta k)} = \underbrace{D \times D \times...\times D}_{\Delta k \hspace{0.1 cm} \mathrm{times}}.
\end{equation}
An explicit form of the elements of this matrix can be inferred for any $\Delta k$ by evaluating the first matrices:
\begin{subequations}
\begin{eqnarray}
D^{(2)} &=& (1-\lambda^2)^2
\left( \begin{array}{ccccc}
1 & 0 & 0 & 0 & \ldots \\
2\lambda^2 & 1 & 0 & 0 & \ldots \\
3\lambda^4 & 2\lambda^2 & 1 & 0 & \ldots\\
4\lambda^6 & 3\lambda^2 & 2\lambda^2 & 1 &  \ldots\\
\vdots & \vdots & \vdots & \vdots & \ddots
\end{array} \right),
\nonumber \\
D^{(3)}&=&(1-\lambda^2)^3
\left( \begin{array}{ccccc}
1 & 0 & 0 & 0 & \ldots \\
3\lambda^2 & 1 & 0 & 0 & \ldots \\
6\lambda^4 & 3\lambda^2 & 1 & 0 & \ldots\\
10\lambda^6 & 6\lambda^2 & 3\lambda^2 & 1 &  \ldots\\
\vdots & \vdots & \vdots & \vdots & \ddots
\end{array} \right),
\nonumber \\
D^{(4)}&=&(1-\lambda^2)^4
\left( \begin{array}{ccccc}
1 & 0 & 0 & 0 & \ldots \\
4\lambda^2 & 1 & 0 & 0 & \ldots \\
10\lambda^4 & 4\lambda^2 & 1 & 0 & \ldots\\
20\lambda^6 & 10\lambda^2 & 4\lambda^2 & 1 &  \ldots\\
\vdots & \vdots & \vdots & \vdots & \ddots
\end{array} \right). \nonumber
\end{eqnarray}
\end{subequations}
Hence, all $D^{(\Delta k)}$ matrices have a similar structure, except for the $(1-\lambda^2)^{\Delta k}$ prefactor, and the numbers accompanying the powers of $\lambda^2$ in the columns, which are given by the $\Delta k$th diagonal of the Pascal triangle. It is then fairly simple to prove by induction that the elements of $D^{(\Delta k)}$ are given by
\begin{equation}
 D_{nm}^{(\Delta k)}=(1-\lambda^2)^{\Delta k}\binom{m+\Delta k-1}{\Delta k-1}\lambda^{2(n-m)}H(n-m).
\label{eq:relation2}
\end{equation}

Note that this general majorization relation implies in particular that $\ket{\Psi_\lambda^{(0)}}\succ\ket{\Psi_\lambda^{(k)}} \hspace{0.1 cm} \forall k$, and therefore, among all Fock state inputs, the vacuum state is the one which minimizes the output entanglement of a two-mode squeezer.

\subsubsection{Proof of $\ket{\Psi^{(0)}_{\lambda'}}\succ\ket{\Psi^{(0)}_{\lambda}}$ for $\lambda'<\lambda$}

It is well known that the entanglement of the two-mode squeezed vacuum state monotonically increases with the squeezing parameter $\lambda$. In what follows we prove a stronger result, that a given two-mode squeezed vacuum state majorizes all the two-mode squeezed vacuum states with stronger squeezing.

We seek for a column-stochastic matrix $R(\lambda,\lambda')$ satisfying
\begin{equation}
 \mathbf{p}^{(0)}(\lambda)=R(\lambda,\lambda')\mathbf{p}^{(0)}(\lambda').
 \label{eq:p0Rp0}
\end{equation}
Based on the matrices of the previous sections, we make an ansatz in which $R$ is a lower-triangular matrix whose columns are all built from a vector $\mathbf{r}(\lambda,\lambda')$, that is,
\begin{equation}
R=\left( \begin{array}{ccccc}
r_0 & 0 & 0 & 0 & \ldots \\
r_1 & r_0 & 0 & 0 & \ldots \\
r_2 & r_1 & r_0 & 0 & \ldots\\
r_3 & r_2 & r_1 & r_0 &  \ldots\\
\vdots & \vdots & \vdots & \vdots & \ddots
\end{array} \right).
\label{Rmat}
\end{equation}
Introducing this ansatz into equation (\ref{eq:p0Rp0}), and recalling that $p^{(0)}_n(x)=(1-x^2)x^{2n}$, we get the following set of linear algebraic equations
\begin{eqnarray}
(1-\lambda^2)&=&(1-\lambda'^2)r_0,
\\
(1-\lambda^2)\lambda^2&=&(1-\lambda'^2)\left(\lambda'^2 r_0+r_1\right), \nonumber
\\
(1-\lambda^2)\lambda^4&=&(1-\lambda'^2)\left(\lambda'^4 r_0+\lambda'^2 r_1+r_2\right), \nonumber
\end{eqnarray}
which can be solved by recursion leading to the solution
\begin{equation}
 r_n=\left(\frac{1-\lambda^2}{1-\lambda'^2}\right)\left[\lambda^2-H(n-1)\lambda'^2\right]
\lambda^{2(n-1)},
\label{eq:r_n}
\end{equation}
which can checked, by induction, to be the solution for a general $n$. Note that $\sum_{n=0}^\infty r_n=1$ as expected.

\subsubsection{Proof of $\ket{\Psi^{(k)}_{\lambda'}}\succ\ket{\Psi^{(k)}_\lambda}$ for $\lambda'<\lambda$}

The same kind of majorization relation can be proved for any $\ket{\Psi^{(k\neq0)}_\lambda}$ state, although the proof is now a little more involved, as we need to find a matrix $R^{(k)}(\lambda,\lambda')$ satisfying
\begin{equation}
 \mathbf{p}^{(k)}(\lambda)=R^{(k)}(\lambda,\lambda')\mathbf{p}^{(k)}(\lambda'),
 \label{eq:pkRpk}
\end{equation}
which now depends on the value of $k$. As we now prove, the  matrix $R^{(k)}(\lambda,\lambda')$ can still be chosen to be lower-triangular, but now every column is defined by its own vector $\mathbf{r}^{(k,j)}$, that is
\begin{equation}
R^{(k)}=\left( \begin{array}{ccccc}
r_0^{(k,0)} & 0 & 0 & 0 & \ldots \\
r_1^{(k,0)} & r_0^{(k,1)} & 0 & 0 & \ldots \\
r_2^{(k,0)} & r_1^{(k,1)} & r_0^{(k,2)} & 0 & \ldots\\
r_3^{(k,0)} & r_2^{(k,1)} & r_1^{(k,2)} & r_0^{(k,3)} &  \ldots\\
\vdots & \vdots & \vdots & \vdots & \ddots
\end{array} \right).
\label{Rk}
\end{equation}

Because we have to recover the case $k=0$ (\ref{eq:r_n}), we make the following ansatz
\begin{eqnarray}
 r_{m}^{(k,n)}&=&\lambda^{2(m-1)}\left(\frac{1-\lambda^2}{1-\lambda'^2}\right)^{k+1} \\
&&\times\left[B_{m}^{(k,n)}\lambda^2-C_{m}^{(k,n)}H(m-1)\lambda'^2\right], \nonumber
\end{eqnarray}
with $B_{m}^{(0,n)}=C_{m}^{(0,n)}=1$, and where the coefficients $B_{m}^{(k\neq0,n)}$ and $C_{m}^{(k\neq0,n)}$ may depend on $\lambda$ and $\lambda'$.

Similarly to the previous section, we can find the coefficients $B_{m}^{(k,n)}$ and $C_{m}^{(k,n)}$ by introducing this ansatz in (\ref{eq:pkRpk}), and using the explicit form of the probability vectors $p_n^{(k)}(x)=(1-x^2)^{k+1}x^{2n}\binom{n+k}{n}$.  Let us show this process step by step.

The system (\ref{eq:pkRpk}) can be rewritten in a compact form as
\begin{equation}
 p^{(k)}_n(\lambda)=\sum_{m=0}^nr^{(k,n-m)}_{m}(\lambda,\lambda')p^{(k)}_{n-m}(\lambda').
\label{eq:relation_R_k}
\end{equation}
For $n=0$, this sets
\begin{equation}
B_{0}^{(k,0)}=1,
\end{equation}
while for $n=1$ we get
\begin{equation}
\lambda^2\binom{k+1}{1}=B^{(k,1)}_{0}\lambda'^2\binom{k+1}{1}+B_{1}^{(k,0)}\lambda^2-C_{1}^{(k,0)}\lambda'^2,
\end{equation}
of which $B_{1}^{(k,0)}=\binom{k+1}{k}$ and $C_{1}^{(k,0)}=B_{0}^{(k,1)}\binom{k+1}{k}$ are valid solutions. Similarly, for $n=2$ (\ref{eq:relation_R_k}) yields
\begin{eqnarray}
&&\lambda^4\binom{k+2}{2}=B_0^{(k,2)}\lambda'^4\binom{k+2}{2}+B_1^{(k,1)}\lambda^2\lambda'^2\binom{k+1}{1} \nonumber
\\
&&-C_1^{(k,1)}\lambda'^4\binom{k+1}{1}+B_2^{(k,0)}\lambda^4-C_2^{(k,0)}\lambda^2\lambda'^2,
\end{eqnarray}
of which $B_2^{(k,2)}=\binom{k+2}{2}$, $C_2^{(k,0)}=B_1^{(k,1)}\binom{k+1}{k}$, and $C_1^{(k,1)}=B_0^{(k,2)}\binom{k+2}{2}/\binom{k+1}{1}$ are now valid solutions.

We observe the pattern of solutions
\begin{subequations}
\begin{eqnarray}
B_m^{(k,0)}&=&\binom{m+k}{k},
\\
C_m^{(k,n)}&=&B_{m-1}^{(k,n+1)}\frac{\binom{n+k+1}{k}}{\binom{n+k}{k}},
\end{eqnarray}
\end{subequations}
so that the components of the vectors $\mathbf{r}^{(k,n)}$ can be rewritten as
\begin{eqnarray}
r_m^{(k,n)}&=&\binom{n+k}{n}^{-1}\left(\frac{1-\lambda^2}{1-\lambda'^2}\right) \label{Newr}
\\
&&\times\left[L_m^{(k,n)}\lambda^2-L_{m-1}^{(k,n+1)}\lambda'^2\right]\lambda^{2(m-1)}, \nonumber
\end{eqnarray}
where we have defined the new parameters
\begin{equation}
 L_m^{(k,n)}=\binom{n+k}{n}B_m^{(k,n)},
\end{equation}
which satisfy $L_m^{(k,0)}=\binom{m+k}{k}$ except for $m<0$, in which case $L_m^{(k,n)}=0$.

In order to find the coefficients $L_m^{(k,n)}$ we use a further condition: as $R^{(k)}(\lambda,\lambda')$ 
must be column-stochastic, the vectors $\mathbf{r}^{(k,n)}$ must be normalized. Let us then define the series
\begin{equation}
S^{(k,n)}=\sum_{m=0}^\infty L_m^{(k,n)}\lambda'^{2m},
\end{equation}
in terms of which the normalization condition $\sum_{m=0}^\infty r_m^{(k,n)}=1$ can be rewritten as
\begin{equation}
\lambda'^2S^{(k,n+1)}=S^{(k,n)}-\binom{n+k}{k}\left(\frac{1-\lambda'^2}{1-\lambda^2}\right)^{k+1}.
\end{equation}
Starting from
\begin{equation}
S^{(k,0)}=\sum_{m=0}^\infty \binom{m+k}{k}\lambda'^{2m}=(1-\lambda'^2)^{-(k+1)},
\label{eq:Sk0}
\end{equation}
these relations allow us to find the rest of $S^{(k,n)}$ recursively, obtaining

\begin{subequations}
\begin{eqnarray}
&&S^{(k,1)}=\lambda'^{-2}(1-\lambda^2)^{-(k+1)}[1-(1-\lambda'^2)^{k+1}],
\\
&&S^{(k,2)}=\lambda'^{-2}(1-\lambda^2)^{-(k+1)}\left\{\lambda'^{-2}\right.
\\
&&\left.-\left[\lambda'^{-2}+\binom{k+1}{k}\right](1-\lambda'^2)^{k+1}\right\}, \nonumber
\\
&&S^{(k,3)}=\lambda'^{-2}(1-\lambda^2)^{-(k+1)}\left\{\lambda'^{-4}\right.
\\
&&\left.-\left[\lambda'^{-4}+\lambda'^{-2}\binom{k+1}{k}+\binom{k+2}{k}\right](1-\lambda'^2)^{k+1}\right\}, \nonumber
\\
&&\quad\vdots \nonumber
\end{eqnarray}
\end{subequations}
from which one sees the general pattern
\begin{eqnarray}
S^{(k,n)}&=&\lambda'^{-2n}(1-\lambda^2)^{-(k+1)} \label{eq:Skn_gen}
\\
&&\times\left[1-(1-\lambda'^2)^{k+1}\sum_{l=0}^{n-1}\lambda'^{2l}\binom{l+k}{k}\right]. \nonumber
\end{eqnarray}
The sum on the right-hand side term can be written in terms of the incomplete beta function
\begin{equation}
B(z;a,b)=\int_0^zdx x^{a-1}(1-x)^{b-1},
\end{equation}
as
\begin{eqnarray}
\sum_{l=0}^{n-1}\binom{l+k}{k}&&\lambda'^{2l}=(1-\lambda'^2)^{-(k+1)}
\\
\times&&\left[1-n\binom{n+k}{k}B(\lambda'^2;n,k+1)\right]. \nonumber
\end{eqnarray}
We can therefore rewrite the condition (\ref{eq:Skn_gen}) as
\begin{eqnarray}
\sum_{m=0}^\infty && L_m^{(k,n+1)}\lambda^{2m}
\\
&&=\lambda'^{-2n}(1-\lambda^2)^{-(k+1)}n\binom{n+k}{k}B(\lambda'^2;n,k+1), \nonumber
\end{eqnarray}
which, given the result (\ref{eq:Sk0}), can be satisfied by choosing
\begin{equation}
L_m^{(k,n)}=n\binom{n+k}{k}\binom{m+k}{k}\lambda'^{-2n}B(\lambda'^2;n,k+1).
\end{equation}
Note that this expression is valid even for $n=0$, as
\begin{equation}
\lim_{a\rightarrow0}aB(x;a,b)=1,
\end{equation}
when $b$ is a positive integer. Introducing this expression for the $L_m^{(k,n)}$ coefficients in $\mathbf{r}^{(k,n)}$ 
(\ref{Newr}), and this into (\ref{Rk}), we get the column-stochastic matrix $R(\lambda,\lambda')$ 
given in the Letter. Hence, we have been able to find a stochastic map connecting 
$\mathbf{p}^{(k)}(\lambda')$ to $\mathbf{p}^{(k)}(\lambda)$, 
which proves the majorization relation $\ket{\Psi^{(k)}_{\lambda'}}\succ\ket{\Psi^{(k)}_{\lambda}}$ if $\lambda'<\lambda$.

\subsection{LOCC protocols}

For completeness, we now give the LOCC protocols corresponding to the previous majorization relations. We believe that these could offer an alternative (more physical) way of attacking the proof of the conjecture for a general input state like (\ref{eq:input_state}), and hence find it appropriate to explain how to build such protocols.

\subsubsection{Transformation $\ket{\Psi_\lambda^{(k+1)}}\rightarrow \ket{\Psi_\lambda^{(k)}}$}

Let us assume that Alice and Bob share the bipartite state $\ket{\Psi_\lambda^{(k+1)}}$, and want to convert it into $\ket{\Psi_\lambda^{(k)}}$. Inspired by the recurrence relation (\ref{eq:relation1}), we propose the following LOCC protocol. Bob starts by performing a POVM measurement \cite{Nielsen-Chuang} described by the measurement operators
\begin{equation}
B_m=\sum_{l=m}^\infty\sqrt{\frac{(1-\lambda^2)\lambda^{2m}p^{(k)}_{l-m}}{p^{(k+1)}_l}}\ket{l-m}\bra{l}.
\end{equation}
Using Eq. (\ref{eq:relation1}), it is easy to verify the condition $\sum_{m=0}^\infty B_m^\dagger B_m=I$.
%$\sum_{m=0}^\infty B_m^\dagger B_m=I_{\{m,\infty\}}$, 
%where $I_{\{m,\infty\}}=\sum_{j=m}^\infty\proj{j}$ is the identity on the support of 
%$\ket{\Psi_\lambda^{(k+1)}}$ on Bob's Hilbert space. 
%Notice that one can easily build a POVM that acts on Bob's full Hilbert space by appending ancillary qubits.
After Bob has completed his local measurement, depending on the outcome $m$ of the measurement, 
the joint state ``collapses" to
\begin{eqnarray}
(I_A\otimes B_m)&\ket{\Psi_\lambda^{(k+1)}}&\propto\sum_{n=m}^\infty \sqrt{p_{n-m}^{(k)}}\ket{n+k+1,n-m} \nonumber
\\
&=&\sum_{n=0}^\infty\sqrt{p_{n}^{(k)}}\ket{n+k+m+1,n}.
\end{eqnarray}
Then, after Bob has communicated the outcome $m$ of his measurement to Alice, 
she performs the local shift operation
\begin{equation}
A_m=\sum_{l=0}^\infty\ket{l}\bra{l+m+1},
\end{equation}
which then yields the desired state $\ket{\Psi_\lambda^{(k)}}$ regardless of $m$, that is, 
deterministically. Remark that the shift operator is trace preserving in the 
subspace spanned by $\{\ket{j+m+1}\}_{j=0,1,..}$, which is the support of $(I_A\otimes B_m)\ket{\Psi_\lambda^{(k+1)}}$ 
on Alice's side. 
Notice that one can easily build a shift operation that acts on Alice's full Hilbert space by appending ancillary qubits.
%Notice that the shift operation can be trivially extended by appending ancillary qubits in order 
%make it act on Alice's full Hilbert space.

\subsubsection{Transformation $\ket{\Psi_\lambda^{(k+\Delta k)}}\rightarrow \ket{\Psi_\lambda^{(k)}}$ for $\Delta k>0$}

Similarly as before but exploiting now (\ref{eq:relation2}), we engineer the following POVM on Bob's side
\begin{equation}
B_m=\sum_{l=m}^\infty \sqrt{\frac{(1-\lambda^2)^{\Delta k} \binom{m+\Delta k-1}{\Delta k-1}\lambda^{2m} p^{(k)}_{l-m}}
{p^{(k+\Delta k)}_l}}\ket{l-m}\bra{l},
\end{equation}
which, combined with the conditional shift in Alice's side
\begin{equation}
A_m=\sum_{l=0}^\infty\ket{l}\bra{l+m+\Delta k},
\end{equation}
deterministically transforms the state $\ket{\Psi_\lambda^{(k+\Delta k)}}$ into $\ket{\Psi_\lambda^{(k)}}$. 
Whenever $k=0$, we obtain the two-mode vacuum squeezed state $\ket{\Psi_\lambda^{(0)}}$, 
which is thus at the end of the majorization chain, and its entanglement is minimum when 
compared to all other states $\ket{\Psi_\lambda^{(k)}}$.

\subsubsection{Transformation $\ket{\Psi^{(0)}_\lambda}\rightarrow \ket{\Psi^{(0)}_{\lambda'}}$ for $\lambda'<\lambda$}

Constructing an LOCC protocol from the stochastic matrix $R(\lambda,\lambda')$ (\ref{Rmat}) 
which connects $\mathbf{p}^{(0)}(\lambda')$ with $\mathbf{p}^{(0)}(\lambda)$ is not an easy task. 
Interestingly, we found a simpler deterministic protocol achieving the same result. 
Let us first give a probabilistic scheme performing the transformation, which we later make deterministic.

\begin{figure}[t]
\begin{center}
\includegraphics[width=\columnwidth]{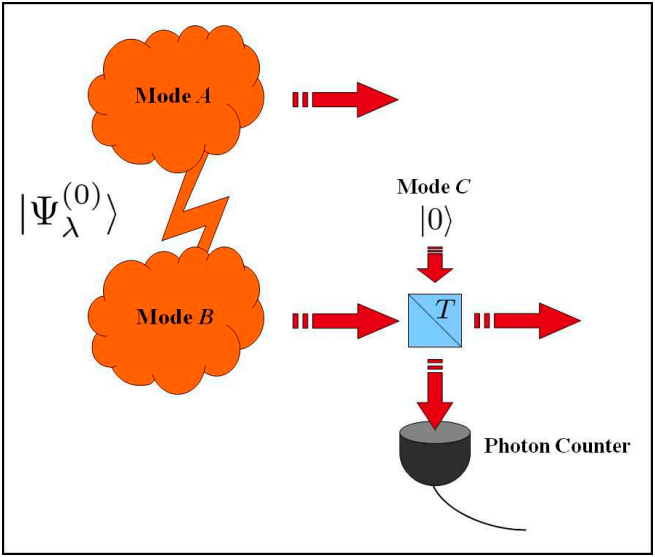}
\end{center}
\caption{Probabilistic LOCC protocol achieving the transformation 
$\ket{\Psi^{(0)}_\lambda}\rightarrow \ket{\Psi^{(0)}_{\lambda'}}$ 
for $\lambda'<\lambda$. Initially, Alice and Bob share the entangled state 
$\ket{\Psi^{(0)}_\lambda}_{AB}$. The first step of the protocol consists 
in Bob mixing his mode $B$ with a vacuum ancillary mode $C$ into a beam-splitter of transmissivity $T$, 
and measuring the number of photons at the output of mode $C$ with a photon counter. 
Conditioned to the measurement of zero reflected photons, the desired transformation is 
achieved with $\lambda'=\sqrt{T}\lambda$.}
\label{Fig:locc_protocol}
\end{figure}

As shown in Figure \ref{Fig:locc_protocol}, Bob mixes his mode $B$ with an ancillary mode $C$ on a beam-splitter 
of transmissivity $T$. The initial state is
\begin{equation}
\ket{\psi}_{ABC}=\ket{\Psi^{(0)}_\lambda}\otimes\ket{0}=\mathcal{N}(\lambda)\sum_{n=0}^\infty\lambda^n\ket{n,n,0},
\end{equation}
where $\mathcal{N}(\lambda)=(1-\lambda^2)^{1/2}$ a normalization factor. After passage through the beam-splitter, 
the joint state becomes
\begin{eqnarray}
\ket{\psi'}_{ABC}=&\mathcal{N}(\lambda)&\sum_{n,m=0}^\infty (T\lambda^2)^{n/2}\left(\frac{1-T}{T}\right)^{m/2} \nonumber
\\
&&\times \binom{n}{m}^{1/2}\ket{n,n-m,m}.
\end{eqnarray}
Finally, Bob measures the number of photons reflected by the beam-splitter. 
The outcome of the measurement will be zero with probability 
$\mathcal{P}=\mathcal{N}^2(\sqrt{T}\lambda)/\mathcal{N}^2(\lambda)$, after which the state will collapse according to
\begin{eqnarray}
\sqrt{\mathcal{P}}\ket{\psi''}_{AB}&=& \left._C\bra{0}\psi'\right\rangle_{ABC}
=\mathcal{N}(\lambda)\sum_{n=0}^\infty T^{n/2}\lambda^n\ket{n,n} \nonumber
\\
&=&\sqrt{\mathcal{P}}\ket{\Psi^{(0)}_{\sqrt{T}\lambda}}_{AB}.
\end{eqnarray}
Then, by choosing the transmissivity of the beam-splitter to satisfy $\lambda'=\sqrt{T}\lambda$ 
we obtain the target state. Note that
there always exists a valid transmissivity $T$, as $\lambda'<\lambda$. 
The input state $\ket{\Psi^{(0)}_\lambda}_{AB}\otimes\ket{0}_C$ being a Gaussian state and the 
projection into vacuum being a Gaussian operation, there must exist a deterministic LOCC protocol 
generating the same outcome \cite{HetVac}. Such a protocol consists of replacing Bob's projection onto vacuum 
by heterodyne detection followed by local displacements on Alice and Bob sides that are proportional to 
the outcome of Bob's heterodyne measurement.

\subsubsection{Transformation $\ket{\Psi^{(k)}_{\lambda}}\rightarrow \ket{\Psi^{(k)}_{\lambda'}}$  for $\lambda'<\lambda$}

Similarly to the case $k=0$, constructing an LOCC protocol from the stochastic matrix 
$R^{(k)}(\lambda,\lambda')$ (\ref{Rk}) which connects
$\mathbf{p}^{(k)}(\lambda')$ with $\mathbf{p}^{(k)}(\lambda)$ is not an easy task. 
Instead, we give a simpler deterministic protocol achieving the same result.

Just as in the previous protocol, Bob starts by mixing mode $B$ with an ancillary mode $C$ on a 
beam-splitter of transmissivity $T$. The joint initial state is
\begin{eqnarray}
\ket{\psi}_{ABC}&=&\ket{\Psi_\lambda^{(k)}}\otimes\ket{0}
\\
&=&\mathcal{N}(k,\lambda)\sum_{n=0}^\infty\lambda^n\binom{n+k}{k}^{1/2}\ket{n+k,n,0}, \nonumber
\end{eqnarray}
with $\mathcal{N}(k,\lambda)=(1-\lambda^2)^{(k+1)/2}$, which becomes
\begin{eqnarray}
&&\ket{\psi'}_{ABC}=\mathcal{N}(k,\lambda)\sum_{n,m=0}^\infty (T\lambda^2)^{n/2}\left(\frac{1-T}{T}\right)^{m/2} \nonumber \\
&&\times\binom{n+k}{k}^{1/2}\binom{n}{m}^{1/2}\ket{n+k,n-m,m},
\end{eqnarray}
after passing through the beam-splitter.

Second, Bob measures the number of photons reflected by the beam-splitter. With probability
\begin{equation}
\mathcal{P}(l)=(1-T)^l\lambda^{2l}\binom{k+l}{l}\frac{\mathcal{N}^2(k,\lambda)}{\mathcal{N}^2(k+l,\sqrt{T}\lambda)},
\end{equation}
the outcome of the measurement will be $l$ photons, and the state of modes $A$ and $B$ will collapse in that case to
\begin{eqnarray}
&\sqrt{\mathcal{P}(l)}&\ket{\psi''}_{AB}= \left._C\bra{l}\psi'\right\rangle_{ABC}
\\
&&=\mathcal{N}(k,\lambda)\left(\frac{1-T}{T}\right)^{l/2} \nonumber
\\
&\times&\sum_{n=l}^\infty(T\lambda^2)^{n/2}\binom{n+k}{k}^{1/2}\binom{n}{l}^{1/2}\ket{n+k,n-l}. \nonumber
\end{eqnarray}
Now, making the variable change $n-l\rightarrow n$ in the sum, and using the relation
\begin{equation}
\binom{n+l+k}{k}\binom{n+l}{l}=\binom{n+k+l}{n}\binom{k+l}{l},
\end{equation}
this state can be rewritten as
\begin{eqnarray}
\sqrt{\mathcal{P}(l)}\ket{\psi''}_{AB}
&=&\mathcal{N}(k,\lambda)(1-T)^{l/2}\lambda^l\binom{k+l}{l}^{1/2}
\nonumber \\
&\times&\sum_{n=0}^\infty(T\lambda^2)^{n/2}\binom{n+k+l}{n}^{1/2}\ket{n+k+l,n} \nonumber \\
&=&\sqrt{\mathcal{P}(l)}\ket{\Psi^{(k+l)}_{\sqrt{T}\lambda}}. 
\end{eqnarray}
Notice that by properly choosing the transmissivity of the beam-splitter so that $\lambda'=\sqrt{T}\lambda$, 
the final state is $\ket{\Psi^{(k+l)}_{\lambda'}}$. Therefore, the last step of the 
protocol consists of applying the transformation $\ket{\Psi^{(k+l)}_{\lambda'}}\rightarrow \ket{\Psi^{(k)}_{\lambda'}}$ 
described above in order to finalize the map $\ket{\Psi^{(k)}_{\lambda}}\rightarrow \ket{\Psi^{(k)}_{\lambda'}}$. 
It is important to remark that our protocol is fully deterministic. Despite the randomness of the photon-counter outcome, 
the determinism is recovered by choosing a different transformation 
$\ket{\Psi^{(k+l)}_{\lambda'}}\rightarrow \ket{\Psi^{(k)}_{\lambda'}}$ 
for each $l$, such that the protocol always ends up in the final state $\ket{\Psi^{(k)}_{\lambda'}}$.


\begin{thebibliography}{99}
\bibitem{footnote} In the definition of Eq.~(\ref{eq:Holevo_info}) the $\sum_a$ should be replaced 
by an integral when considering a continuous alphabet $a$.
\end{thebibliography}
\end{document}